\documentclass[preprint,12pt]{elsarticle}


\usepackage{amssymb}

\usepackage{lineno}

\usepackage{graphicx}
\usepackage{subcaption}
\usepackage{amsmath}
\usepackage{float}
\graphicspath{{Figure/}}
\usepackage{xcolor}
\usepackage{tabularx}
\usepackage{hyperref}
\hypersetup{
    colorlinks=true,
    linkcolor=blue,
    filecolor=blue, 
    citecolor=blue,
    urlcolor=blue,
    pdftitle={Manuscript},
    pdfpagemode=FullScreen,
    }

\begin{document}

\begin{frontmatter}



\title{An Interfacial Profile-Preserving Approach for Phase Field Modeling of Incompressible Two-Phase Flows}

\author[inst1]{Haohao Hao}
\author[inst1]{Xiangwei Li}
\author[inst1]{Chenglin Jiang}
\author[inst1]{Huanshu Tan\corref{cor1}}
\ead{tanhs@sustech.edu.cn}
\cortext[cor1]{Corresponding author}

\affiliation[inst1]{organization={Multicomponent Fluids group, Center for Complex Flows and Soft Matter Research \& Department of Mechanics and Aerospace Engineering, Southern University of Science and Technology},
            city={Shenzhen},
            postcode={518055}, 
            state={Guangdong},
            country={China}}



\begin{abstract}

In this paper, we introduce an interfacial profile-preserving approach for phase field modeling for simulating incompressible two-phase flows. While the advective Cahn-Hilliard equation effectively captures the topological evolution of complex interfacial structures, it tends to displace the fluid interface from its equilibrium state, impacting simulation accuracy. 
To tackle this challenge, we present an interfacial profile-preserving formulation that relies on a phase-field-related signed distance function, rather than the phase field function itself.
It is solved iteratively to restore the equilibrium interface profile after each time step.
This approach effectively minimizes discretization errors and enhances mass conservation accuracy for each phase. 
Our formulation is discretized using a second-order Total Variation Diminishing (TVD) Runge-Kutta method within iterations and a finite volume scheme in spatial discretization.
We quantitatively compare our present profile-preserving method with the original method in terms of accuracy and convergence rate through simulations of a deforming drop in a single vortex and a rising bubble in quiescent fluid, and further validate the applicability through simulations of a two-dimensional contracting liquid filament, a drop impacting a deep liquid pool, and three-dimensional drop deformation in shear flow. 
Our results exhibit good agreement with analytical solutions, prior numerical results, and experimental data, demonstrating the effectiveness and accuracy of our proposed approach.

\end{abstract}


\begin{highlights}
\item  An improved conservative phase ﬁeld method is developed for simulation of the two-phase incompressible flows.

\item A Profile-preservation formation is proposed to keep the interfacial profile at equilibrium state.

\item Mass conservation errors for each phase have been much reduced.

\item Applicability of proposed method is demonstrated by comparing present results with prior numerical results, theoretical, and our experimental data.
\end{highlights}


\end{frontmatter}


\section{Introduction}\label{sec:s1}

Two-phase flows are prevalent in natural phenomena and find numerous applications in engineering. 
Over the past decades, various methods have been proposed to model two-phase flows, including volume of fluid~\cite{hirt1981volume}, front tracking~\cite{unverdi1992front}, Level set~\cite{sussman1999adaptive} and phase field methods~\cite{anderson1998diffuse,jacqmin1999calculation}.
Among these, the phase field method has received increasing attention and achieved remarkable success in simulating two-phase flow~\cite{jacqmin1999calculation,badalassi2003computation,yue2004diffuse,magaletti2013sharp,guo2021well, zhang2023central}. 
Compared to other methods, the phase field method possesses several important advantages, including its ability to follow the topological evolution of complex structures without explicit tracking, flexibility to extend and implement for multi-phase flow simulation~\cite{kim2012phase,mao2021variational}, and features of maintaining ~\textit{total} mass conservation~\cite{yue2007spontaneous} and having the order parameter physically associated with free energy functional~\cite{cahn1958free,jacqmin1999calculation,yue2004diffuse}. 
Therefore, phase field method has demonstrated successful applications in simulating a wide range of two-phase flow processes, including but not limited to two-phase turbulent flows~\cite{scarbolo2015coalescence,LIU2021110659}, phase changes~\cite{badillo2012quantitative,yang2022abrupt}, flows of non-Newtonian fluids\cite{yue2006phase}, dynamics of contact lines~\cite{jacqmin2000contact}, and droplet dynamics within micro-channels~\cite{de2006modeling,carlson2010droplet}.

To accurately simulate multiphase flow, required are the preservation of interfacial profile and its accurate discretization, as phase field method is a representative diffuse (thick) interface method.
Phase field methods usually apply an advective Cahn-Hilliard equation to govern the deformation of the diffuse interface.
However, the interplay between convection and diffusion terms can cause the thickening and thinning of the fluid interface, thereby the deviation from the required hyperbolic tangent profile (equilibrium state of the diffuse interface)~\cite{jacqmin1999calculation}.
Discretization error of the interface in space and time can cause numerical diffusion, and the fluid interface slight deviates from its equilibrium state.
These deviations inevitably affect the accuracy of surface tension implementation at the interface and the representation of fluid properties, generating uncertainties to the simulated results,
particularly in modeling surface-tension related hydrodynamics. 
Despite the fact that reasonable choice of mobility or peclet number can mediate the convection-diffusion interplay and alleviate this issue to an extent~\cite{jacqmin1999calculation,ding2007diffuse}, there are no general principles or standards for its optimal value selection.
This limitation affects the applicability of the advective Cahn-Hilliard equation.

Some artificial phenomena have been observed and reported in the phase field simulation work, such as shrinkage and coarsening of static drops or bubbles\cite{yue2007spontaneous,dai2014coarsening} due to normal gradient of the chemical potential \cite{zhang2017flux}. 
This results into the unphysical dissolution and mass non-conservation for each phase~\cite{zhang2010cahn}. \citet{yue2007spontaneous} have given the analytical solution of drop shrinkage in the absence of convection based on the exchange between interfacial and bulk energies. For the coarsening effect, smaller domains will be diminished as nearby larger domains expand\cite{zhang2017flux}, even in the sharp interface limit of the Cahn-Hilliard equation \cite{dai2014coarsening,lee2016sharp}.  
Addressing these undesired effects typically involves increasing the mesh resolution and decreasing the time step size~\cite{yue2007spontaneous,ding2007diffuse}. However, such approaches significantly escalate computational costs and reduce the efficiency of the phase field method.


To overcome the aforementioned drawbacks of the original phase field while minimizing computational costs, several researchers have proposed modified Cahn-Hilliard equations with various correction terms in the recent years, such as profile correction~\cite{li2016phase}, flux correction~\cite{zhang2017flux}, interface compression~\cite{zhang2019interface} and mass correction~\cite{wang2015mass}. 
Subsequently, the modified Cahn-Hilliard equations with the profile correction term and the flux correction term was applied in simulating turbulent multiphase flows \cite{soligo2019mass}.
The basic idea of these corrections is to keep equilibrium profile of the interface by introducing a penalty flux to counteract the chemical-potential-gradients-induced flux, or by re-distributing mass sources or sinks in diffuse interface to guarantee the mass conservation of each phase.

While these methods have shown improvements in terms of mass conservation and interface profile, the physical interpretation of the correction or compression term in the modified Cahn-Hilliard equations remains unclear, leading to uncertainty regarding its influence on interface convection in the simulation~\cite{li2016phase,soligo2019mass,zhang2019interface}. 
Furthermore, the behavior of the correction term depends on the chosen Peclet number~\cite{li2016phase,zhang2017flux,soligo2019mass}, which adds complexity to the application of these methods in simulating incompressible two-phase flows.

In this paper, inspired by the conservative level set method \cite{olsson2005conservative} that facilitates the development of the conservative Allen–Cahn model for multiphase flow \cite{chiu2011conservative,chiu2019coupled}, we will introduce an interfacial profile-preserving approach by maintaining the equilibrium state of the diffuse interface during time iteration, instead of modifying the the original Cahn-Hilliard equation as reported in previous work~\cite{li2016phase,zhang2017flux,soligo2019mass,zhang2019interface}.
A modified artificial profile preservation equation, based on the relationship between the order parameter and the signed distance function~\cite{waclawczyk2015consistent}, will be proposed to reduce the interface dispersion and consequently improve the accuracy of the surface tension and fluid properties modeling.
Through this preserving approach, effect of the choice of the peclet number in the advective Cahn-Hilliard equation can be less pronounced and mass conservation errors for each phase can be reduced.

We organize this paper as follows: Section \ref{sec:s2} describes the numerical methodology, including the governing equation, numerical discretization, analysis of truncation error and solution procedure. 
In Section \ref{sec:s3}, to assess the mesh convergence and accuracy of our proposed phase field method, we present a drop deforming in a vortex and 2D axisymmetric bubble rising.
Section \ref{sec:s4} provides numerical simulations, including contracting liquid filament, drop deformation in a shear flow and impact of a drop on a deep liquid pool, to extensively demonstrate the capabilities of our approach.
In this section, we also compare our results with numerical, theoretical and experimental results in previous studies. 
Finally, we conclude the paper in Section \ref{sec:s5}. 

\section{Numerical model}\label{sec:s2}

\subsection{Phase field method formulation}

The interface is represented by the order parameter or the phase field function \(c(\mathbf{x},t)\) and implicitly tracked by the advective Cahn-Hilliard equation\cite{ding2007diffuse}

\begin{equation} \label{211}
\ \frac{\partial c }{\partial t}+ \nabla \cdot (\mathbf{u} c) = \frac{1}{Pe} \nabla^2 \psi ,\ 
\end{equation}	

\begin{equation} \label{212}
\ \psi=\xi^{\prime}(c)-Cn^2 \nabla^2 c,
\end{equation}  
where \(\mathbf{u}\) is the  dimensionless fluid velocity, \(Pe=(M^*/Cn)^{-1}\) is Peclet number which represents the ratio between convection and diffusion for the order parameter \(c\). \(M^*\) is the mobility number.  \(Cn=\alpha/L\) is the Cahn number, denoting the dimensionless interface thickness, where \(\alpha\) is the interface thickness and \(L\) is the characteristic length of the investigated system. Unless otherwise stated, \(Cn\) is set to \(0.5 \Delta x\), where \( \Delta x\) is the grid size. 
Building upon the selection suggested by Magaletti et al. [8], we employ the relations \(M^*\thicksim Cn^2\) and \(Pe \thicksim Cn^{-1}\), demonstrating that the diffuse interface converges toward the sharp interface as \(Cn\) approaches zero.
The chemical potential \(\psi\) comprises two distinct terms.  
The first term originates from the bulk energy within a two-phase fluid system and is defined by the a double well function \(\xi(c)= c^2(1-c)^2/4\).  
Meanwhile, the second term serves as a correction introduced for address the influence of a two-phase interface within the system.


\subsection{Profile-preservation formulation}

The key issue of the original phase field in Eq.\eqref{211} is the deviation of phase field function from the hyperbolic tangent function, leading to inaccurate representation of surface tension and fluid properties. In this section, we present the formulation of a profile-preserving equation designed to maintain the hyperbolic tangent profile of the diffuse interface.


An intermediate step in conservative level set method \cite{olsson2005conservative} has been proposed to keep the profile of the level set function constant, which is similar to the reinitialization of original level set method \cite{sussman1999adaptive}.
The conservation form of this step can be written as \cite{olsson2007conservative}
\begin{equation} \label{221}
\frac{\partial c }{\partial \tau }=\nabla \cdot \{[\epsilon (\nabla c \cdot \mathbf{n})-c(1-c)]\mathbf{n}\},
\end{equation}	
where \(\tau\) is an artificial time and \(\epsilon\) is a small parameter proportional to the dimensionless interface thickness. The first term \(\epsilon (\nabla c \cdot \mathbf{n})\mathbf{n}\) on the right-hand side is a small amount of viscosity to decrease the stationary shocks with \(\tau\) increasing. The other term \(c(1-c)\mathbf{n}\) on the right-hand side of this equation corresponds to the compressive flux \cite{olsson2007conservative}. \(\mathbf{n}\) is the normal direction of the interface and calculated by

\begin{equation} \label{228}
\mathbf{n(c)}=\frac{\nabla c_{\tau=0}}{\left|\nabla c\right|_{\tau=0}}.
\end{equation}

When the interface reaches the equilibrium, the chemical potential \(\psi\) is constant throughout the domain. Using Equation~\eqref{212}, we can determine the equilibrium profile \(c(\mathbf{x})\) by solving \(\Delta \psi = 0\), which results in the expression~\cite{ding2007diffuse}
\begin{equation} \label{2221}
c(\mathbf{x})=0.5(1+\tanh(\frac{\phi(\mathbf{x})}{2 \sqrt{2} Cn})). 
\end{equation}
In this equation,  \(\phi\) represents the signed distance function,  defined as the distance from point \(\mathbf{x}\) to the nearest point on the interface \(\Gamma\). 
Consequently, \(c\) undergoes a gradual transition between the two bulk values (\(c=0 \) and 1). 
An algebraic relation between \(c\) and \(\phi\) is further elucidated in reference~\cite{waclawczyk2015consistent}, i.e.,
\begin{equation}\label{222}
\phi=\sqrt{2} Cn \ln (\frac{c}{1-c}).
\end{equation}

After substituting the equilibrium profile function Eqn.~\eqref{2221} into the compressive flux term in Eq.\eqref{221}, we can obtain
\begin{equation} \label{223}
{c}{(1-c)}=\frac{1}{4}(1-{{\tanh }^{2}}(\frac{\phi }{2 \sqrt{2} Cn })).
\end{equation}	
While the left and right terms in the above expression are mathematically equal, opting to utilize the signed distance function \(\phi\), as opposed to \(c\), for computing the non-linear compressive flux proves effective in reducing discretization errors. This advantage arises from the fact that the second derivative of \(\phi\) is zero, whereas that of \(c\) is nonzero. A more detailed explanation will be provided in Section \ref{sec:s241} and \ref{sec:s242}.



\begin{figure}[H]
    \centering
    \includegraphics[width=0.9\textwidth]{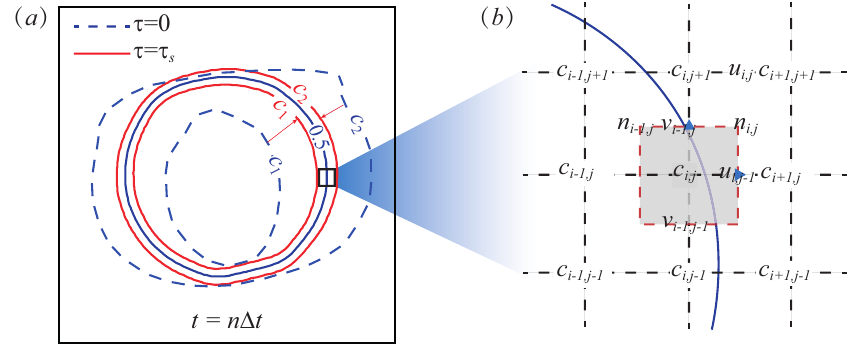}
    \caption{(a) Schematic of profile-preservation formulation. (b) Configuration of the position of the phase field function, velocity and interface normal. The grey block denotes the control volume. Note that only some variables and 
 symbols are shown to clearly illustrate the MAC mesh.}
    \label{fig:mesh241}
    \end{figure}

The gradient of the hyperbolic tangent function \(c\) and the signed distance function \(\phi\) have the relation derived by \citet{waclawczyk2015consistent}, i.e.,
\begin{equation} \label{224}
\nabla c=\frac{{c}{(1-c)}}{\sqrt{2} Cn} \nabla \phi.
\end{equation}
Thus, the normal direction of the interface can also be obtained by
\begin{equation} \label{225}
\mathbf{n}=\frac{\nabla c}{\left|\nabla c\right|} =\frac{\nabla \phi}{\left|\nabla \phi\right|}.
\end{equation}
After substituting Eq.~\eqref{228}, Eq.~\eqref{223} and Eq.~\eqref{225} into Eq.~\eqref{221}, we get a profile-preserving equation, written as
\begin{equation} \label{226}
\frac{\partial c }{\partial \tau }=\nabla \cdot \{[\epsilon (\nabla c \cdot \mathbf{n(\phi)})-\frac{1}{4}(1-{{\tanh }^{2}}(\frac{\phi }{2 \sqrt{2} Cn }))]\mathbf{n(\phi)}\},
\end{equation}	
where \(\mathbf{n}=\nabla \phi_{\tau=0} /\left|\nabla \phi\right|_{\tau=0}\) and \(\epsilon = \sqrt{2}Cn\). In this equation, the calculation of the compressive flux term relies on the signed distance function \(\phi\) rather than the phase field function \(c\). 
This approach effectively minimizes discretization errors.   
To preserve the equilibrium profile of the fluid interface (\(0<c<1\)) during each iterative computation over time \(t\) for the advective Cahn-Hilliard equation, as illustrated in Fig.~\ref{fig:mesh241}(a), we iteratively solve this equation with respect to \(\tau\) until it satisfies the steady state criteria defined by

\begin{equation}\label{227}
\int_{\Omega}\left|c_n^{m+1}-c_n^m\right| d \Omega \leq T O L \cdot \Delta \tau.
\end{equation}

 The subscript notation n is n-\(th\) time step and corresponds to the time \(t = n \Delta t\), the superscript notation m-\(th\) artificial correction step, and \(T O L\) is the threshold value. After iterating the Eq.\eqref{226} to steady state, the interface profile will be restored to its equilibrium. In practice, this iteration typically takes only several time steps. Unless specified otherwise, we choose \(\Delta \tau = 0.01\Delta x\) and \(T O L = 1\).

In the implementation, a small parameter \(\zeta\) is introduced to avoid \(\phi\) approaching \(\infty\) or \(-\infty\) when \(c\) goes to 0 or 1. Hence, the modified relation can be expressed as
\begin{equation}\label{2271}
\phi = \sqrt{2} Cn \ln \left(\frac{c+\zeta}{1-c+\zeta}\right),
\end{equation}
where the value of \(\zeta\) determines the region of \(\phi\) around the fluid interface. 
It need to be wider than the interface thickness to accurately compute the flux in Eq.\eqref{223} and the curvature in Eq.\eqref{233}. In present work, \(\zeta\) is set to \(\zeta=10^{-16}\). It should be noted that \(\phi\) is only an approximation to a signed-distance function in Eq.\eqref{222}.

\subsection{Navier–Stokes equation}

The two-phase flows can be considered as incompressible, governed by the dimensionless Navier–Stokes equations with surface tension and continuity equation. It can be written as 
\begin{equation} \label{231}
\ \nabla \cdot \textbf{u} = 0,\
\end{equation}
\begin{equation} \label{232}
\ \rho(\frac{\partial \textbf{u}}{\partial t}+\textbf{u}\cdot \nabla\textbf{u})=-\nabla p+\frac{1}{Re}\nabla \cdot [\mu(\nabla\textbf{u}+\nabla\textbf{u}^T)]+\frac{\boldsymbol{f_s}}{We}+ \frac{\rho}{Fr}\textbf{j},\
\end{equation}
where \(\rho\), \(p\), \(\mu\), \(\boldsymbol{f_s}\) and \(\boldsymbol{j}\) represent density, pressure, viscosity, surface tension, and the direction of gravity, respectively. The fluid density and viscosity depend on the phase field function and are given by \(\rho(c)=c \rho_{1}+(1-c) \rho_{2}\) and \(\mu(c)=c \mu_{1}+(1-c) \mu_{2}\). The subscript 1 and 2 denote the fluid 1 and fluid 2, respectively. Additionally, the dimensionless numbers are the Reynolds number  \(Re=\rho_1 U D/\mu_1\), the Weber number \(We=\rho_1 U^2D/\sigma\), and the Froude number \(Fr=U^2/{gD}\), where \(\sigma\), \(g\), and \(D\) represent the surface tension coefficient, the gravitational acceleration, and the characteristic length, respectively.


The surface tension force \(\boldsymbol{f_s}\) we use in present study is approximated as

\begin{equation}\label{233}
\boldsymbol{f_s} = -\kappa(\phi) \nabla c,
\end{equation}
where the curvature \(\kappa\) is calculated by the signed distance function \(\phi\), as \(\kappa(\phi) =\nabla \cdot \frac{\nabla \phi}{\left|\nabla \phi\right|}\). Similar to the Eq.\eqref{226}, it is also more accurate to calculate curvature using the well-behaved signed distance function, compared to the previous surface tension formulation where the curvature calculated using the phase field function \cite{kim2005continuous,lee2012regularized}.




\subsection{Numerical method} \label{sec:s24}

\subsubsection{Discretization of Cahn-Hilliard and NS equation } \label{sec:s240}

The momentum equation Eq.\eqref{231} is solved using the finite-volume method based on the second order accurate central difference schemes on a rectangular uniform marker-and-cell (MAC) mesh. The velocity vector field {\bf u} is defined at cell faces, the interface normal \(\mathbf{n}\) is defined at cell vertex and the scalar ﬁelds (\(p\), \(\mu\), \(\rho\), \(c\) and \(\phi\)) are deﬁned at cell centers, as shown in Fig.\ref{fig:mesh241}(b). The advection term and viscous term in Eq.\eqref{231} are discretized using the second-order Adams–Bashforth scheme and Crank–Nicolson scheme in time, and using the centered second-order scheme in space, respectively. The advection term in Eq.\eqref{211}, is discretized using a fifth-order weighted essentially non-oscillatory (WENO) scheme, where the local ﬂow velocity determines the up-winding direction. The standard projection method is used to solve the coupled equations Eqs.\eqref{231} and \eqref{232}, where the contributions of surface tension and gravity to the momentum in Eq.\eqref{231} are taken into account when the intermediate velocity is computed at the first step. The Poisson equation is solved using an efficient multigrid method with the Gauss–Siedel method as a smoother. More details can be found in Ref.\cite{spelt2005level} and \cite{ding2007diffuse}.

\subsubsection{Discretization of the interfacial profile-preserving equation } \label{sec:s241}

The second order TVD Runge-Kutta method \cite{gottlieb1998total} and the finite volume scheme are applied to discretize the profile-preserving equation Eq.\eqref{226} in the temporal and spatial dimension. Here, we take the discretisation of the Eq.\eqref{226} in two dimensions as an example and they are relatively straightforward to implement in three dimensions. The Eq.\eqref{226} can be approximated in two-dimension space as

\begin{equation} \label{241}
\begin{aligned}
& \frac{\partial c_{i,j}}{\partial \tau}=\left\{\frac{F_{i+1/2,j}-F_{i-1/2,j}}{\Delta x}+\frac{F_{i,j+1/2}-F_{i,j-1/2}}{\Delta y}\right\}-\\
& \quad \quad \quad \left\{\frac{G_{i+1/2,j}-G_{i-1/2,j}}{\Delta x}+\frac{G_{i,j+1/2}-G_{i,j-1/2}}{\Delta y}\right\},\\
&
\end{aligned}
\end{equation}
where \(F\) and \(G\) are the numerical flux and defined at cell centers, the subscript \(i+1/2\) and \(j+1/2\) denote the variables defined at cell faces. These fluxes are evaluated by
\begin{equation}\label{242}
\begin{aligned}
& {F}_{i \pm 1/2,j}= \left. \epsilon n_x\left(\frac{\partial c}{\partial x} n_x+\frac{\partial c}{\partial y} n_y\right)\right|  _{i \pm 1/2,j},\\
& {F}_{i,j\pm 1/2}=\left. \epsilon n_y \left(\frac{\partial c}{\partial x} n_x+\frac{\partial c}{\partial y} n_y\right)\right|_{i,j\pm 1/2},\\
& {G}_{i \pm 1/2,j}=  \frac{n_x}{4}\left[1-\tanh ^2\left(\frac{{\phi}}{2 \sqrt{2} \epsilon}\right)\right]  _{i \pm 1/2,j},\\
& {G}_{i ,j\pm 1/2}=  \frac{n_y}{4}\left[1-\tanh ^2\left(\frac{{\phi}}{2 \sqrt{2} \epsilon}\right)\right]  _{i,j \pm 1/2}.\\
&
\end{aligned}
\end{equation}

The interface normal in Eqs.\eqref{226} and \eqref{233} is calculated using the standard central difference approximation, as
\begin{equation}\label{243}
\begin{aligned}
& \left. n_x\right|  _{i,j}=\left. \frac{\frac{\partial \phi}{\partial x}}{\sqrt{|\frac{\partial \phi}{\partial x}|^2+|\frac{\partial \phi}{\partial y}|^2}}\right|_{i,j}, 
 \left.\quad n_y\right|  _{i,j}=\left. \frac{\frac{\partial \phi}{\partial y}}{\sqrt{|\frac{\partial \phi}{\partial x}|^2+|\frac{\partial \phi}{\partial y}|^2}}\right|  _{i,j},\\
&
\end{aligned}
\end{equation}
where the partial derivative of \(\phi\) is calculated using the standard central difference approximation, as
\begin{equation}\label{2435}
\begin{aligned}
& \left. \frac{\partial \phi}{\partial x}\right|  _{i,j}=\frac{ \phi_{i+1,j+1/2}-\phi_{i,j+1/2}}{2 \Delta x}, \left. \frac{\partial \phi}{\partial y}\right|  _{i,j}=\frac{ \phi_{i+1/2,j+1}-\phi_{i+1/2,j}}{2 \Delta y}.\\
&
\end{aligned}
\end{equation}

For variable solved at cell face (\(i \pm 1/2,j\) or \(i,j \pm 1/2\) ) in above equation, we can approximate its value by second-order linear interpolation. For example, if \(F\) and \(\phi\) are solved at \((i+1/2,j)\), then
\begin{equation}\label{2432}
\begin{aligned}
& {F}_{i+1/2,j} = 0.5({{F}_{i+1,j}}+{{F}_{i,j}}),\\
& {\phi}_{i+1/2,j} = 0.5({{\phi}_{i+1,j}}+{{\phi}_{i,j}}).\\
&
\end{aligned}
\end{equation}

\subsubsection{Truncation error}\label{sec:s242}

 In attempt to understand the practical difference in the truncation error due to linear interpolation in Eq.\eqref{2432} between the signed distance function \(\phi(\mathbf{x})\) and phase field function (hyperbolic tangent function, \(c(\mathbf{x})\)), therefore, for simplicity, we assume these two function \(\phi\) and \(c\) are one-dimension. These can be expressed as
\begin{equation}
 \phi(x)=x, \quad c(x)=0.5(1+\tanh({\phi(x)}/(2\sqrt{2} Cn))).
 \end{equation}
 
  The position of the interface is given at \(x=0\). The Taylor expansion of \({{\phi }_{i}}\) and \({{\phi }_{i+1}}\) at \((i+1/2)\) can be expressed as
\begin{equation} \label{2441}
\begin{aligned}
  & {{\phi }_{i+1}}={{\phi }_{i+1/2}}+{{{{\phi }'}}_{i+1/2}} (0.5\Delta x) +{{{{{\phi }''}}}_{i+1/2}} {{(0.5\Delta x)}^{2}}/2!+\\
  &   \quad \quad \quad {{{{{{\phi }'''}}}}_{i+1/2}} {{(0.5\Delta x)}^{3}}/3!+{\mathrm O}(\Delta {{x}^{4}}), \\ 
 & {{\phi }_{i}}={{\phi }_{i+1/2}}-{{{{\phi }'}}_{i+1/2}} (0.5\Delta x)+{{{{{\phi }''}}}_{i+1/2}} {{(0.5\Delta x)}^{2}}/2!-\\
 &\quad \quad \quad {{{{{{\phi }'''}}}}_{i+1/2}}{{(0.5\Delta x)}^{3}}/3!+{\mathrm O}(\Delta {{x}^{4}}), \\ 
&
\end{aligned}
\end{equation}
where the superscript \(\prime\) and the exclamation mark \(!\) denote the derivative of \(\phi\) and the factorial, respectively. The discretization error \(E_{\phi}\) can be obtained by
\begin{equation}\label{2442}
\begin{aligned}
  & E_{\phi}=0.5({{\phi }_{i+1}}+{{\phi }_{i}})-{{\phi }_{i+1/2}}={{{{{\phi }''}}}_{i+1/2}}{{(0.5\Delta x)}^{2}}+{\mathrm O}(\Delta {{x}^{4}}). \\ 
&
\end{aligned}
\end{equation}

 Similarly, we also express the discretization error \(E_{c}\) as
\begin{equation}\label{2443}
\begin{aligned}
 & E_{c}=0.5({{c}_{i+1}}+{{c}_{i}})-{{c}_{i+1/2}}={{{{c}''}}_{i+1/2}} {{(0.5\Delta x)}^{2}}+{\mathrm O}(\Delta {{x}^{4}}), \\ 
&
\end{aligned}
\end{equation}
where the Taylor expansion of \(c(x)\) and \(c''(x)\) at \(x=0\) can be expressed as
\begin{equation} \label{2444}
\begin{aligned}
  & c(x)=\frac{1}{2}(1+\frac{x}{2\sqrt{2}Cn}-\frac{1}{3}{{(\frac{x}{2\sqrt{2}Cn})}^{3}}+\frac{2}{15}{{(\frac{x}{2\sqrt{2}Cn})}^{5}}+{\mathrm O}({{x}^{7}})), \\ 
 & {c}''(x)=\frac{1}{2}(-\frac{1}{4C{{n}^{2}}}(\frac{x}{2\sqrt{2}Cn})+\frac{1}{3C{{n}^{2}}}{{(\frac{x}{2\sqrt{2}Cn})}^{3}}+{\mathrm O}({{x}^{5}})). \\ 
\end{aligned}
\end{equation}

We define the fluid interface (diffuse interface) to be in the range \(0.05\leq c(x)\leq 0.95\), where \(x\) varies from \(-4.164Cn\) to \(4.164Cn\). 
It becomes evident that \({\phi''(x) }\) equals zero within the diffuse interface region, while \({c''(x) }\) is nonzero except at the interface \((x=0)\).
Consequently, the discretization error of \({c(x)}\) is greater than that of \({\phi }\) at same mesh size \(\Delta x\).

\subsubsection{Solution procedure}

\begin{figure}[H]
\centering
\includegraphics[width=0.5\textwidth]{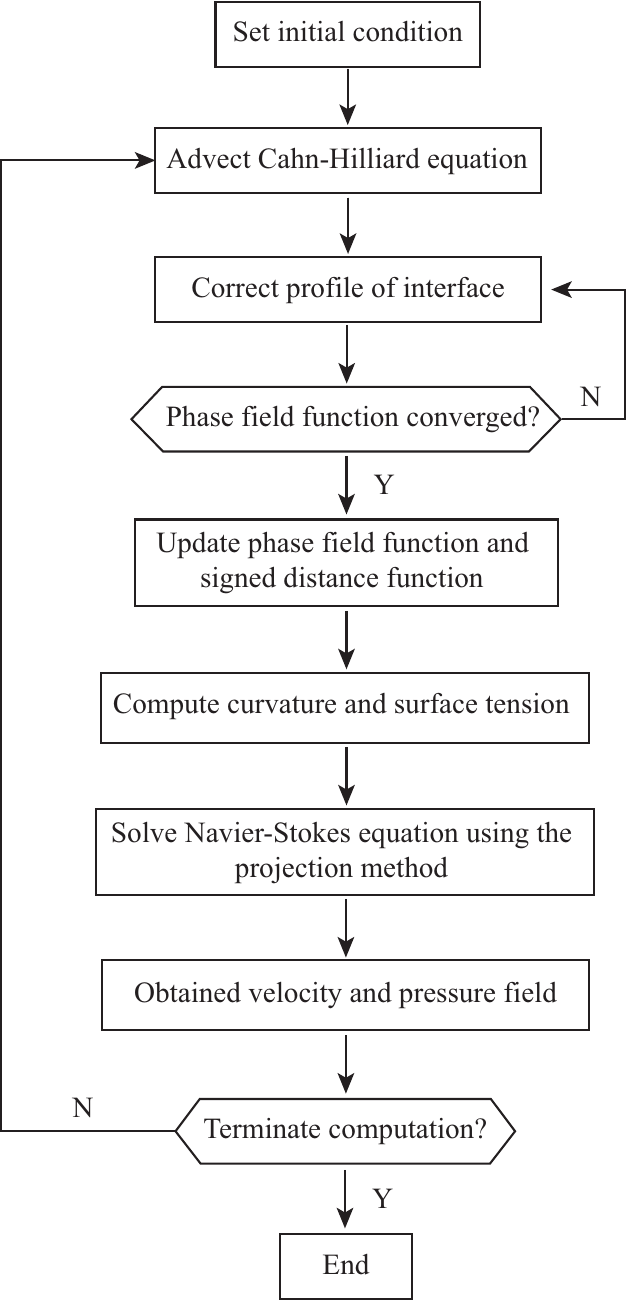}
\caption{Flow chart of the computational procedure}
\label{fig:mesh242}
\end{figure}

As shown in Fig. \ref{fig:mesh242}, the overall solution procedure for one time step loop can be summarized as follows
\begin{enumerate}
\item Advect the phase field function \(c(\mathbf{x},t)\) using Eq.\eqref{211} after initialization. 
\item Correct the profile of the phase field \(c(\mathbf{x},t)\) using Eq.\eqref{226} and determine whether a steady state criteria has been reached using Eq.\eqref{227}.
\item Update the phase field \(c(\mathbf{x},t)\) and compute the signed distance function \(\phi(\mathbf{x},t)\) using Eq.\eqref{2271}.
\item Compute the curvature and surface tension using Eq.\eqref{233}.
\item Solve the Navier-Stokes equation and continuity equation using the projection method.         
\item Obtain the new velocity and pressure field.

\end{enumerate}    

\section{Phase field transport tests}\label{sec:s3}

In this section, two 2D simulations consisting of a drop deforming in a single vertex and a rising bubble in quiescent fluid, are performed to evaluate and analyse the performance of the present phase field. To quantify the improvement of the proposed phase field method in terms of mass conservation of each phase, we define the error \(E_m\) of mass as

\begin{equation} \label{321}
E_m = E(t) - E_{i}=\int_{\Omega }{\left| {{c }}-{{c }_{i}} \right|}dV,
\end{equation}
where the subscript \(i\) denotes the initial moment.

\subsection{Drop deformation}\label{sec:s31}

\begin{figure}[H]
\centering
\includegraphics[width=1.0\textwidth]{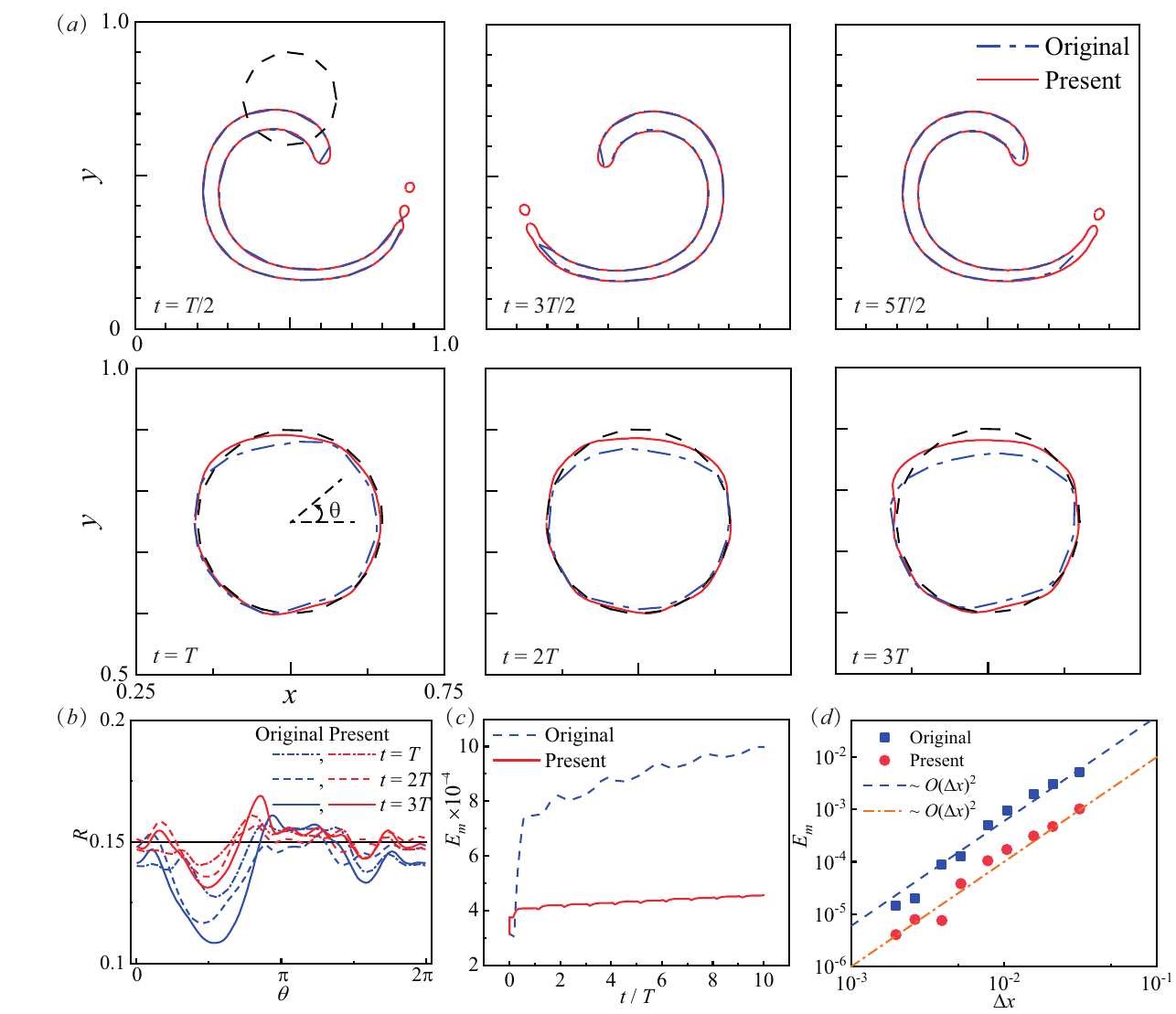}
\caption{(a) Interface shape of the drop (\(c=0.5\)) in a single vortex with a grid size of \(D/\Delta x=128\) at different times. The black dash line denotes the exact solution. The red and blue lines represent the results with and without the profile correction. (b) Contour of \(c=0.5\) in polar coordinate system at \(t=T, 2T\) and \(3T\). (c) Comparison of the temporal evolution of the mass ratio \(E/E_{in}\) at \(D/\Delta x=128\) for the original and present phase field method. (d) Convergence of the mass error \(E_m\) at \(t = 2T\) for the original and present phase field method.}
\label{fig:mesh311}
\end{figure}

The deformation of a drop in a single vortex, as proposed by \citet{rider1998reconstructing}, has been widely utilized to assess the effectiveness of various methods \cite{desjardins2008accurate,chiu2019coupled} in capturing  interfaces. 
The shearing velocity is described as follows,
\begin{equation} \label{311}
 u=-{{\sin }^{2}}(\pi x/D)\sin (2\pi y/D)\cos \left( \frac{\pi t}{T} \right), 
\end{equation}
\begin{equation} \label{312}
  v=\sin (2\pi x/D){{\sin }^{2}}(\pi y/D)\cos \left( \frac{\pi t}{T} \right). 
\end{equation}
Here, the time period \(T\) is set to 4. 
As illustrated in Fig.~\ref{fig:mesh311}, an initially circular drop (indicated by the black dashed line) with a radius of \(R=0.15\) is positioned at \((x_0,y_0)=(0.5,0.75)\) within a unit square domain.
Notably, the drop reaches its maximum deformation state at odd times of \(T/2\), while returning to its initial shape at even multiples of \(T/2\) in theory.

Fig.~\ref{fig:mesh311}(a) shows the deformation of drop with a grid size of \(D/\Delta x=128\) and the Peclect number of \(Pe=1/Cn\) at \(t=T/2, T, 3T/2, 2T, 5T/2\), and \(3T\). We can see that the present phase field method (indicated by the red solid line) have more powerful capability of capturing the interface profile than the original phase field (the blue dashed line) by comparing the captured interface with the exact solution. To further highlight the improved interface topology by the present method, Fig.\ref{fig:mesh311}(b) displays the contours of \(c=0.5\) at \(t=T, 2T\), and \(3T\) in polar coordinate system. It can be found that compared with the original method, the interfaces obtained by the present method deviate less from the exact solution.
 
As mentioned in Section \ref{sec:s1}, the original phase field method is efficient tool for simulating two-phase flow since the total amount of \(c\) is conserved globally. However, it does not necessarily conserve mass for each fluid \cite{zhang2010cahn}. Hence, the mass conservation of a drop is examined at moderate resolution. Fig.\ref{fig:mesh311} (c) presents the comparisons of the mass error \(E_m\) against time with the mesh size of \(D/\Delta x=128\). 
It shows that the \(E_m\) increases very slowly with the exception of the initial stage and the drop mass is always maintained to a high degree of accuracy for the present method, while it increases faster for the original phase field method due to numerical dissipation and diffusion term in Cahn-Hilliard equation \cite{wang2015mass,zheng2014shrinkage,soligo2019mass}. 

To evaluate the convergence of the present method, we select various test cases with different grid resolutions (\(D/\Delta x=32, 48, 64, 96, 128, 192, 256, 384\), and 512). 
For all cases, the time step is set to \(\Delta t=4 \times 10^{-4}\), and the mass error \(E_m\) is plotted against the mesh size in Fig.\ref{fig:mesh311} (d). 
It is found that the \(E_m\) for the present method is typically an order of magnitude less than those of the original phase field at the same resolution, although the convergence rates are approximately two for these two methods. 
    
\begin{figure}[H]
\centering
\includegraphics[width=1.0\textwidth]{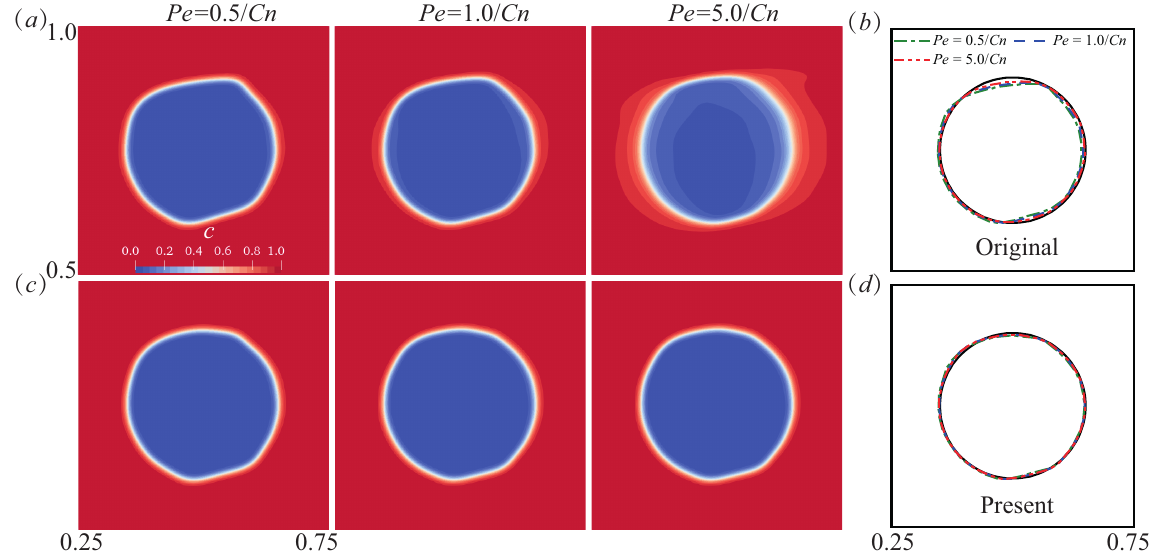}
\caption{Comparison of the instantaneous profile of the phase field function and interface shape of the drop with three different \(Pe\) numbers at \(t=T\) and \(D/\Delta x=192\) for the original (upper) and present (lower) phase field method.}
\label{fig:mesh314}
\end{figure} 

To illustrate the influence of the numerical \(Pe\) number in the advective Cahn-Hilliard equation, we present the instantaneous profile of the phase field function and the drop interface shape for both the original and present phase field methods in  Fig.~\ref{fig:mesh314}.  
As \(Pe\) increases (depicted in Figs.~\ref{fig:mesh314}a and b), the original phase method exhibits improved accuracy in recovering the initial shape of the drop.
However, it is important to note that the phase field function \(c\) deviates significantly from its equilibrium state in this case.
This deviation occurs when the advective Cahn-Hilliard equation is dominated by the convection term (a large \(Pe\)), and the diffusion term struggles to restore the interface to equilibrium.
In contrast, for the proposed phase field approach ( Figs.~\ref{fig:mesh314}c and d), the finial shapes of the drop remain consistent across different \(Pe\) values.
Consequently, the dependence of the phase field function \(c\)  on the choice of \(Pe\) number is minimized. 
Simultaneously, fluid convection does not significantly thicken the fluid interface, which remains approximately in an equilibrium state.

\subsection{Axisymmetric rising bubble}\label{sec:s32}

\begin{figure}[h!]
\centering
\includegraphics[width=0.4\textwidth]{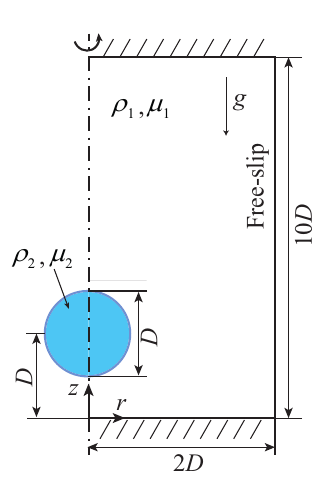}
\caption{Schematic of a rising bubble}
\label{fig:mesh321}
\end{figure}

The simulation of an axisymmetric bubble rising are carried out to check the performance of the profile-corrected formulation in Eq.\eqref{226} and the surface tension model in Eq.\eqref{233}. 
The schematic of this problem is shown in Fig. \ref{fig:mesh321}. A bubble with diameter \(D\) is initially placed at \(z=D\) in a rectangular domain of \(2D \times 10D\). To increase the time step \(\Delta t\) and decrease the computational cost, the density and viscosity ratios are chosen as \(\rho_2/\rho_1=0.01\) and \(\mu_2/\mu_1=0.01\), respectively. Wall conditions are used at the upper and lower boundaries. Symmetric boundary condition and free-slip boundary condition are applied at the left and right boundaries, respectively. In addition, we choose \(D\) as the the characteristic length and \(U=\sqrt{gD}\) as the characteristic velocity. The dimensionless parameters are defined as \(Re=\rho_1 U D/\mu_1\), \(We=\rho_1 U^2D/\sigma\), and \(Fr=U^2/{gD}\), set to \(Re=35\), \(We=100\), and \(Fr=1\), respectively.

\begin{figure}[h]
\centering
\includegraphics[width=0.9\textwidth]{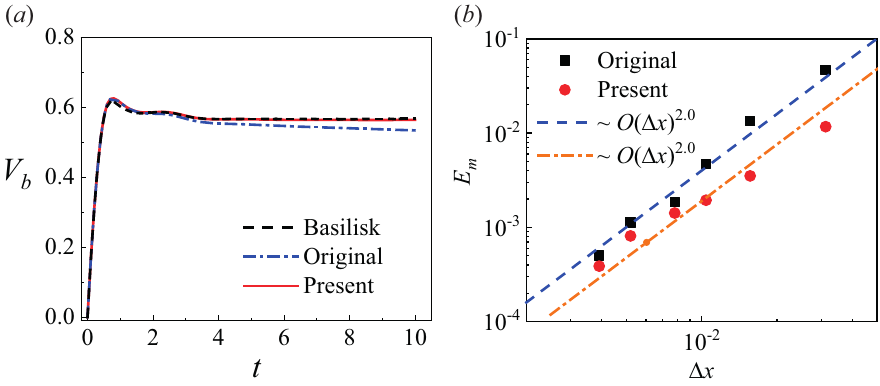}
\caption{(a) Evolution of the rising velocity of bubble at \(D/\Delta x=128\). (b) Convergence of the mass error \(E_m\) at \(t = 7\) for the original and present phase field method.}
\label{fig:mesh322}
\end{figure}

To show the accuracy of the present phase field method, the rising velocity \(V_b\) of bubble against time are quantitatively compared with the results obtained by the original phase field method and the open-source multiphase flow solver (Basilisk) \cite{popinet2009accurate} where the geometric volume of fluid (VOF) method has been implemented with the same resolution. Fig.\ref{fig:mesh322}(a) shows that the rising velocity obtained by the present phase field has excellent agreement with those of the geometric VOF. Moreover, the convergence rates of the mass error \(E_m\) using Eq. \eqref{321} are also examined on different grids (\(D/\Delta x=32\), 64, 96, 128, 192 and 256). The time step size is \(\Delta t = 5 \times 10^{-4}\) for all cases except that \(\Delta t = 2 \times 10^{-4}\) for the mesh size of \(D/\Delta x=256\). In Fig.\ref{fig:mesh322}(b), the mass error \(E_m\) could be reduced by an order of magnitude in comparison to the original phase field, especially at lower and moderate resolution. The convergence rates for both two methods are approximately 2.

\begin{figure}[H]
\centering
\includegraphics[width=1.0\textwidth]{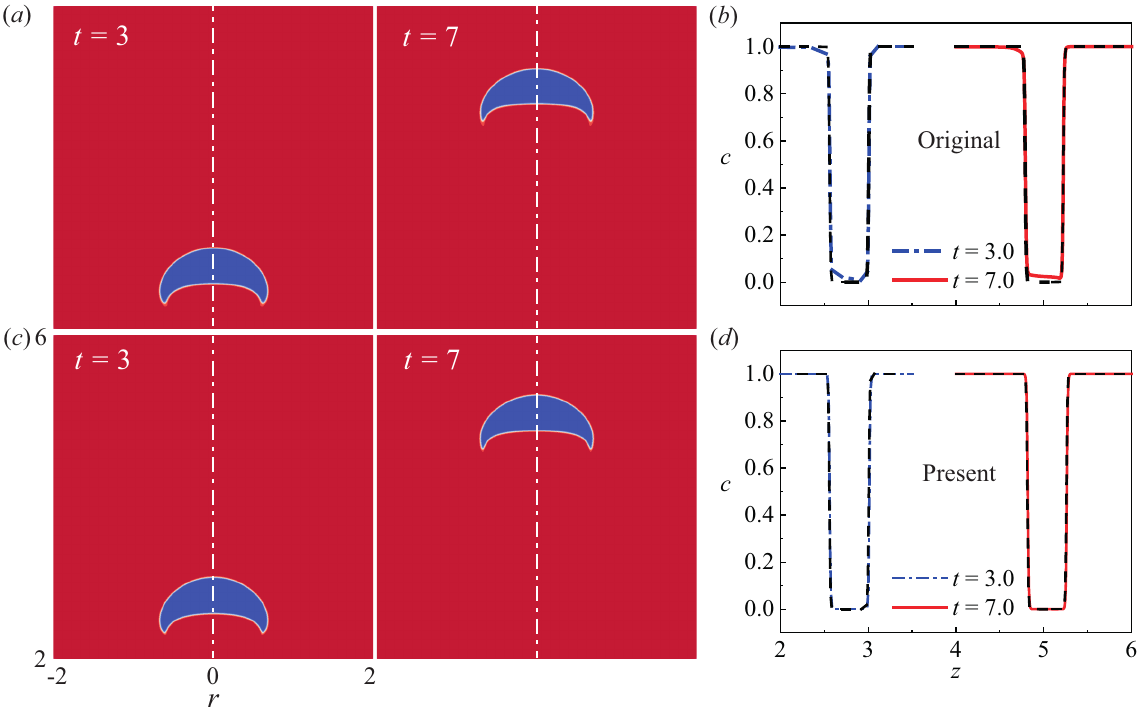}
\caption{Comparison of the interface shape of the rising bubble and the instantaneous profile of the phase field function \(c\) along the dash-dotted line (\(r = 0\)) at \(D/\Delta x=128\) (Upper: original, Lower: present). The black dashed lines in (b) and (d) represent the theoretical equilibrium profile.}
\label{fig:mesh323}
\end{figure}

In Fig.\ref{fig:mesh322}(a), the bubble reaches a steady state for the results of the present phase field and the geometric volume of fluid (VOF) method. This phenomenon was also observed by the experiment by \citet{hnat1976spherical} and the numerical simulation by \citet{chiu2019coupled}. 
However, it can be clearly observed that the the terminal rising velocity obtained by the original phase field method is smaller than the result of the present phase field method and its difference continues to grow over time. The reason for this is that inaccurate implementation of the surface tension and accumulation of mass errors of bubble, since the interface profiles deviates from its equilibrium and its extent tends to increase as the bubble rises, as shown in Fig.\ref{fig:mesh323}(a). Compared to those calculated from original phase field method, the interface profiles obtained through the proposed approach are well preserved and always keep its equilibrium in present phase field method in Fig.\ref{fig:mesh323}(b).

\section{Applications}\label{sec:s4}

To demonstrate the applicability of proposed approach in capturing complex interface deformation, involving pinch-off phenomena, we carried out two numerical experiments: one involving a 2D axisymmetric contracting liquid filament and another concerning the 3D deformation of a drop in a shear flow. Followed by these simulations, we present a comprehensive comparison of flow field and interfacial evolution in the context of a drop impacting a deep liquid pool, juxtaposed with our experimental results. 
Unless otherwise specified, we choose \(Pe = 1.0/Cn\) for numerical simulations.


\subsection{Contracting liquid filament}\label{sec:s41}

First, we consider the liquid filament contracting under the action of surface tension \cite{stone1986experimental} to validate the present phase field method coupled with Navier–Stokes equation. The retraction dynamics of the filament depends on the Ohnesorge number (\(Oh = \mu/\sqrt{\rho \sigma R}\)), the aspect ratio (\(\Gamma = L/R\)) and the initial surface perturbation \cite{notz2004dynamics,driessen2013stability}. It may pinches off and collapses into multiple droplets, which is mainly due to the Rayleigh–Plateau instability for viscous filaments (\(Oh >0.1\)) \cite{driessen2013stability} and the end-pinching mechanism for low-viscosity filaments (\(Oh \leq 0.1\)) \cite{stone1986experimental,notz2004dynamics,anthony2019dynamics}.

\begin{figure}[H]
\centering
\includegraphics[width=0.18\textwidth]{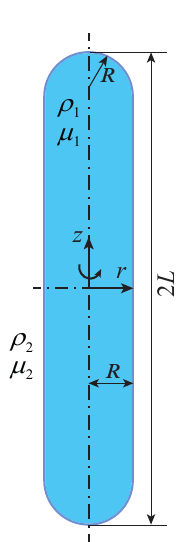}
\caption{Schematic of contracting liquid filament}
\label{fig:mesh411}
\end{figure}

The filament we study here consists of an axisymmetic cylinder and the hemispherical caps at its both ends. Its total length is \(2L\), and the radii of axisymmetic cylinder and hemispherical caps are \(R\), as depicted in Fig.~\ref{fig:mesh411}. The radius \(R\) and the inertial-capillary time $\sqrt{\rho R^3\sigma}$ are chosen as the characteristic length and time, respectively. Therefore, the Weber number is \(We=1\) and the Reynolds number \(Re\) can be considered as the  reciprocal of the Ohnesorge number. In present simulation, the dimensionless parameters are \(Oh=0.01\), \(\Gamma=15\), the density ratio \(\rho_2/\rho_1=0.001\) and the viscosity ratio \(\mu_2/\mu_1=0.01\). Considering the axisymmetic filament is symmetric at \(z=0\) and \(r=0\), we therefore simulate only a quarter of the filament. In addition, The computational domain has a size of \(5R \times 20R\), which is large enough to avoid the effect due to the domain size. The mesh size and time step size are \(R/\Delta x=64\) and \(\Delta t =5 \times 10^{-4}\), respectively. Symmetric boundary is used at the boundaries of \(z=0\) and \(r=0\), and the Neumann boundary condition is applied at the other two boundaries. 

\begin{figure}[H]
\centering
\includegraphics[width=1.0\textwidth]{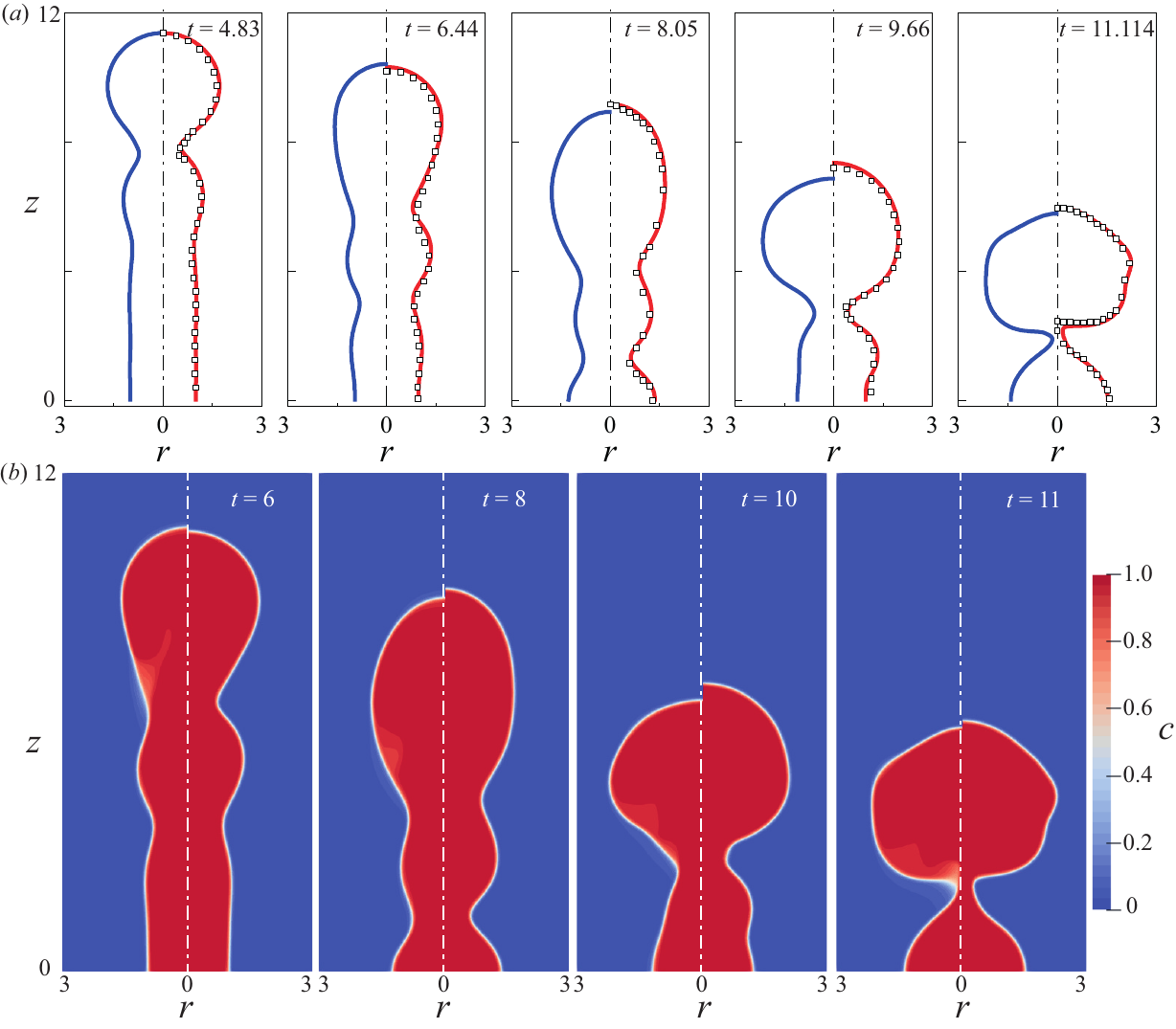}
\caption{(a) Instantaneous shapes of a contracting filament at \(\Gamma=15\) and \(Oh=0.01\). The blue and red lines represent the results obtained by the original (left) and present (right) phase field, respectively. The black squares denote the results from the previous numerical results\cite{notz2004dynamics}.(b) Instantaneous contours of the phase field function \(c\) for the original (left) and present (right) phase field at \(Oh=0.01\) and \(\Gamma=15\). Only a quarter of the filament is shown.}
\label{fig:mesh412}
\end{figure}

We begin by comparing our present results with numerical results (depicted as square dots) calculated using the sharp interface method by~\citet{notz2004dynamics}. In Figure~\ref{fig:mesh412}(a), the blue (left) and red (right) lines represent the results obtained using the original and present phase field methods (diffuse interface), respectively.
It is obvious that there is good agreement between the sharp interface method and our proposed approach.
Fig.\ref{fig:mesh412}(b) shows the instantaneous profile of the phase field function \(c\) obtained by the original and present phase field method at different times. 
We can see that for the original method, the interfaces in the vicinity of the neck have been thickened due to strong convection and large gradient of velocity around this region. As a result, the balance between the convection term and diffusion term in Eq.\eqref{211} has been broken, which results in the interface profile deviating from its equilibrium and thus inaccurate implementation of the surface tension. These errors could be continuously accumulated and finally the pinch-off behaviour of the filament could not be captured naturally. 

\subsection{Three-dimensional droplet in a shear flow}\label{sec:s42}    

The 3D drop deformation in a shear flow are also known to be a good example for the phase field method applicable to the simulation of two-phase flows to test its accuracy and the efficiency \cite{badalassi2003computation,yue2004diffuse}. The dimensionless parameters controlling its deformation dynamics include the capillary number \(Ca= \mu k R/\sigma\), the Reynolds number \(Re= \rho k R^2/\mu\), and the viscosity ratio \cite{taylor1932viscosity,badalassi2003computation}, where \(k\) is the shear rate and \(R\) is spherical drop radius. The Taylor deformation parameter \(D_{T}\) can describe the deformation of drop, defined as 

\begin{equation} \label{421}
\ D_{T}=\frac{L_T-B_T}{L_T+B_T}\
\end{equation}
where \(L_T\) and \(B_T\) denotes the longest and shortest axes of drop in the shear plane when its shape reaches the steady state. For equal viscosity, the relation between the \(D_{T}\) and the \(Ca\) was first established by \citet{taylor1932viscosity} at \(Ca \ll 1\) and \(Re \ll 1\), written as \(D_T=(35/32)Ca\). It means that the shear stress exerted by the surrounding fluid is physically balanced by the surface tension force at equilibrium.

\begin{figure}[H]
\centering
\includegraphics[width=0.6\textwidth]{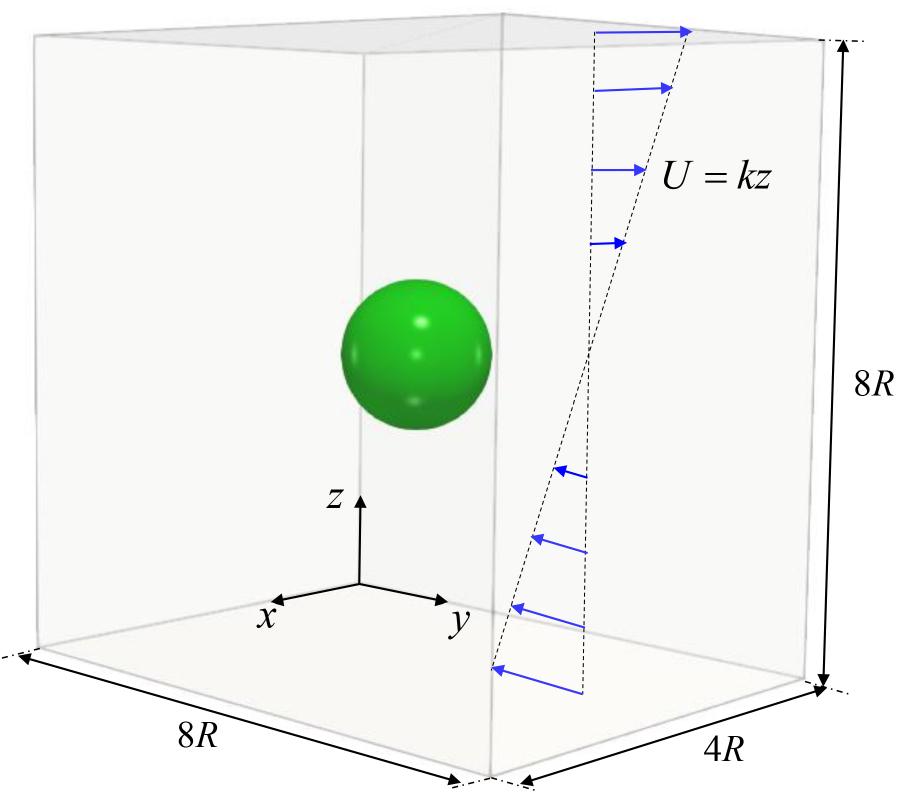}
\caption{Schematic of an initially spherical droplet in a shear flow}
\label{fig:mesh421}
\end{figure}

As illustrated in Fig.\ref{fig:mesh421}, the spherical drop with a radius of \(R\) is initially placed at the center of the fluid domain. The simulations are performed on a computational domain of size \(4R \times 8R \times 8R\) in shear flow with the velocity \(\textbf{u}=(0,kz,0)\), which has demonstrated that the numerical results are independent of its size. \(R\) and \(1/k\) are chosen as the characteristic length and time and thus the characteristic velocity equals \(kR\). The Reynolds number is chosen as 0.08. The mesh size and time step are \(D/dx=16\) and \(dt = 5\times 10^{-5}\), respectively. Here, we set the density and viscosity ratios to 1 to compare the present results with the analytical solution \cite{taylor1932viscosity}. In addition, No-slip boundary condition is applied in \(z\) direction while symmetric boundary condition and periodic boundary condition are imposed in \(x\) and \(y\) direction, respectively. 

\begin{figure}[H]
\centering
\includegraphics[width=1.0\textwidth]{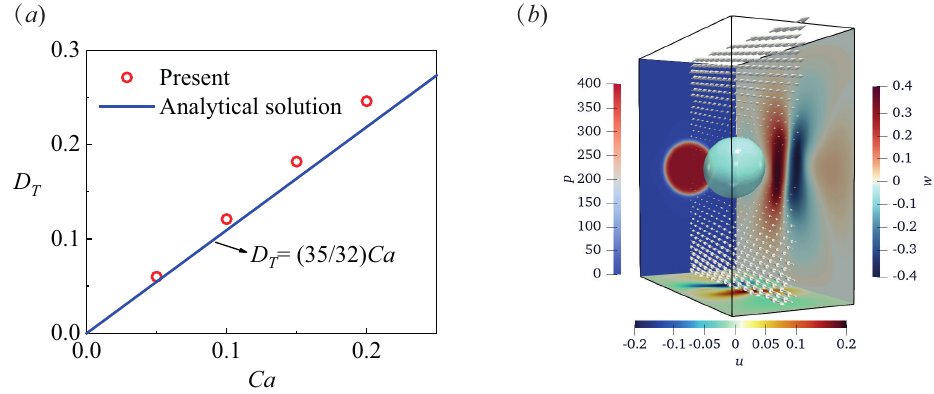}
\caption{(a) Comparison of the present results with the analytical solution \cite{taylor1932viscosity} in terms of Tayer deformation parameter \(D_{T}\). (b) Equilibrium state of a droplet deforming in a shear flow and the velocity vector in middle section (\(x=0\)) at \(Ca = 0.05\) and \(Re = 0.08\). Note that we translate the pressure field at \(y=4R\), the velocity field in \(x\) and \(z\) directions at \(z=4R\) and \(x=2R\) to the computational boundaries for better visualization purpose.} 
\label{fig:mesh422}
\end{figure}

In Fig.\ref{fig:mesh422}(a), it can be seen that the present results agrees well with the analytical solution for \(Ca \leq 0.1\). However, the difference between these results increases when \(Ca \geq 0.15\) since the assumption of \(Ca \ll 1\) can not be satisfied. When the drop reaches an equilibrium state, its shape becomes ellipsoidal and the pressure difference between the inside and the outside of the drop is approximately the Laplace pressure (\(2\sigma/R\)) at its initial spherical shape, as shown in Fig.\ref{fig:mesh422}(b).

\subsection{Drop impact on a deep liquid pool }\label{sec:s43}


To further demonstrate the application of the proposed method, we will give a quantitative comparison between our numerical and experimental results for the impact of a drop on a deep pool. Its impact processes involve interface coalescence and crater formation, which is affected by many factors, such as the drop diameter, impact velocity and physical properties of the fluids (density, viscosity, and surface tension) \cite{prosperetti1993impact,murphy2015splash}.

The experimental setup is similar to those used in recent experimental studies \cite{lherm2023velocity}, where the drop impact velocity \(U\) changes with its initial height \(H_i\). The 100 cSt silicone oil was used both in the drop of diameter \(D=1.74\) mm and in the pool. Its density and viscosity are \(\rho_1 = 960\) kg/m\(^3\) and \(\mu_1 = 96\) mPa\(\cdot\)s. The surface tension of the air–oil interface is \(\sigma=20.1\) mN/m. The ratios of density and viscosity are \(\rho_2 / \rho_1=1.28 \times 10^{-3}\) and \(\mu_2 / \mu_1= 1.88 \times 10^{-5}\). It is noted that multigrid solvers for poisson equation in Eq.~\eqref{232} have difficulty converging for very high density and viscosity ratios \cite{tryggvason2011direct}. Considering that the exact density and viscosity ratios play a relatively minor role in dynamics of drop impact when these are large enough, we set \(\rho_2 / \rho_1=1 \times 10^{-2}\) and \(\mu_2 / \mu_1= 1 \times 10^{-2}\) in simulation to speed up the calculation.

\begin{figure}[H]
\centering
\includegraphics[width=0.5\textwidth]{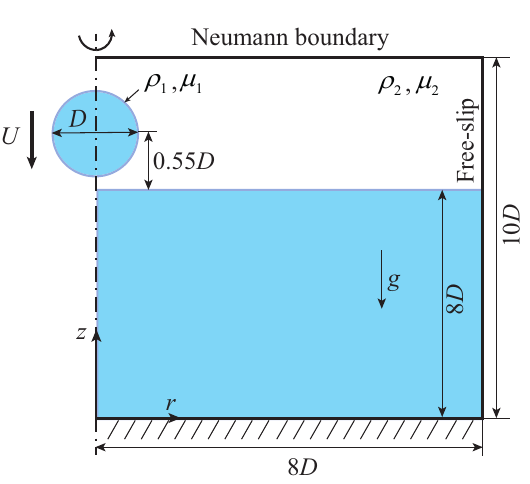}
\caption{Schematic of a drop impacting on the liquid interface}
\label{fig:mesh431}
\end{figure}

The schematic of the numerical setup is shown in Fig.\ref{fig:mesh431}. The computation is performed in 2D axisymmetric domain of \(8D \times 10D\), which can avoid the reflection of the capillary wave from the wall. Initially, the height of liquid surface and drop are \(8D\) and \(8.55D\), respectively. Symmetric boundary condition and no-slip boundary condition are enforced at the left boundary and other three boundaries, respectively. 

\begin{figure}[H]
\centering
\includegraphics[width=1.0\textwidth]{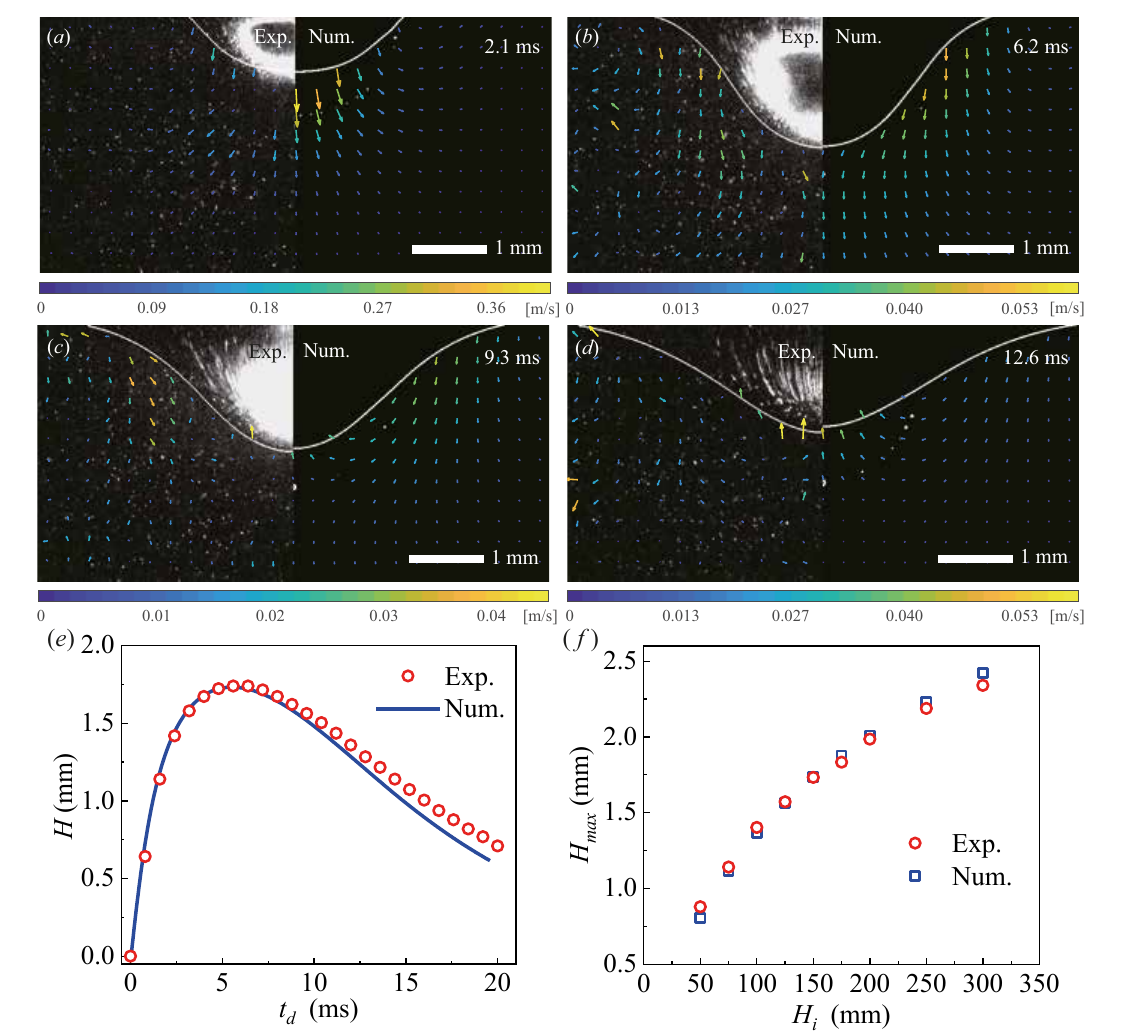}
\caption{Comparison of a drop impact onto a deep liquid pool between experiments and numerical simulations. The snapshots of the velocity vector field are at \(t=\) 2.1 (a), 6.2 (b), 9.3 (c) and 12.6 ms (d) at the initial drop height \(H_i\) = 150 mm. \(t=0\) ms corresponds to the time when the drop reaches the liquid surface. The while lines represent the air-oil interface. (e) Evolution of the depth \(H\) of the crater at \(H_i\) = 150 mm. \(t_d=0\) ms corresponds to the time when the shape of the crater is observed. (f) The maximum crater depths \(H_{max}\) as a function of \(H_i\). }
\label{fig:mesh432}
\end{figure}

 A quantitative comparison of a drop impact on a deep liquid pool between experiment and simulation are presented in Fig.\ref{fig:mesh432}. The left and right side of Fig.\ref{fig:mesh432} (a-d) show the experimentally and numerically obtained crater profile and velocity vector field, respectively. It is can be seen that the simulated crater profile and the velocity field are in good agreement with experiment results, especially in the expansion stage of the crater. This can also be observed from the evolution of the crater depth in Fig.\ref{fig:mesh432} (e). Besides, we give the comparison of the maximum crater depth with different \(H_i\) and a fair agreement can be observed again, as shown in Fig.\ref{fig:mesh432} (f). Thus, the applicability and accuracy of the present phase field method are confirmed.

\section{Conclusion}\label{sec:s5}

In summary, this study has yielded an improved conservative phase field method for simulating the incompressible two-phase flows. 
Our approach introduces a modified profile-preservation formulation to address the challenge of non-equilibrium interfacial profiles during the iterative process of advective Cahn-Hilliard equation. 
Notably, this formulation applies a well-behaved signed distance function to compute the compressive term, significantly reducing truncation errors.
The second order TVD Runge-Kutta method is used for time discretization and the finite volume method is used for the spatial discretization. 
The primary advantage of this method is its ease of implementation and reduced dependence on numerical \(Pe\) numbers, all while incurring minimal computational overhead.

Through numerical simulations, including scenarios such as a drop in a single vortex, 2D axisymmetric rising bubbles, axisymmetric contracting liquid filaments, 3D drop deformation in a simple shear flow, and drop impact on a deep pool, we have successfully validated the effectiveness and accuracy of the present method.
The results align closely with other numerical simulations, theoretical solutions, and experimental data, showcasing the method's capability to excel, particularly in scenarios where surface tension effects dominate.

Looking ahead, our future research endeavors will focus on extending this profile-preserving formulation to simulations of N-phase flows, offering broader opportunities for exploration and application within the realm of intricate fluid dynamics scenarios.

\section*{Acknowledgements}
This work has received financial support from the Natural Science Foundation of China (Grant No. 12102171) and the Natural Science Foundation of Shenzhen (Grant No. 20220814180959001).

\appendix


 \bibliographystyle{elsarticle-num-names} 
 \bibliography{cas-refs}





\end{document}